# Thermal equilibration between excited states or solvent effects: unveiling the origins of anomalous emissions in heteroleptic Ru(II) complexes

Denis Jacquemin,[a] Daniel Escudero,[a,*]

In this manuscript we present a computational study on the photoluminescence properties of several heteroleptic [Ru(H)(CO)(N^N)(tpp)$_2$]$^+$ complexes (tpp=triphenylphosphine). A special focus is set on disentangling the temperature-dependent emissive properties. Experimentally, when cooling down a solution of [Ru(H)(CO)(dmphen)(tpp)$_2$]$^+$ (dmphen=5,6-dimethyl-1,10-phenanthroline) from room to cryogenic temperatures, a partial emission switch from a metal-to-ligand charge transfer ($^3$MLCT) to a ligand-centered ($^3$LC) phosphorescence is observed, resulting in dual photoluminescence. Two different origins of the anomalous emissions are possible, i.e., thermal equilibration between electronically excited states or different excited state solvent relaxation effects. Our calculations are in favor of the thermally-equilibrated scenario. This computational investigation highlights the importance of controlling the temperature-dependent emissive behavior for optoelectronic applications.

Controlling the temperature-dependent photoluminescent properties is of vital importance in optoelectronics applications and more in particular in the field of organic light emitting diodes (OLEDs).[1] Since the seminal work of Adachi on the use thermally activated delayed fluorescence (TADF)[2] dyes as an efficient strategy to overcome the limited electroluminescent efficiencies of purely organic dyes in OLEDs devices, increasing efforts have been devoted to the understanding of thermally-mediated photoluminescent processes. Indeed, the TADF mechanism roots on the use of thermal energy to assist reverse intersystem crossing from the singlet to the triplet excited states (ES). Thus, delayed fluorescence is only observed if enough thermal vibrational energy is available to overcome the singlet-triplet splitting. TADF is not found only organic dyes but has also been reported in coordination complexes of, e.g., Cu[3] and Ag;[4] where it often co-exists with prompt phosphorescence. In these complex photochemical scenarios a subtle competition between multiple ES deactivation processes is at play; and only the fastest processes are being observed, i.e., there is kinetic control.[5] Anti-Kasha fluorescence or phosphorescence,[6] i.e., emissions from higher-lying ES, are among these complex photochemical scenarios. Two different anti-Kasha cases exist: a thermally-equilibrated scenario and a non-thermally equilibrated one.[7] The latter one is characterized by: i) large ES$_n$-ES$_1$ (where ES can be a singlet or a triplet state) energy gaps, so that the sluggish ES$_n$→ES$_1$ internal conversion (IC) and/or intersystem crossing (ISC) processes cannot compete with a photon emission from the higher-lying ES; and ii) excitation wavelength-dependent photoluminescence properties. Belonging to non-thermally equilibrated scenarios, dual singlet-triplet emitters of, e.g., Os(II)[8], Ag(I) and Pt(II)[9] as well as dual triplet-triplet emitters of, e.g., Ru(II)[10] and Ir(III)[11] have been reported in the literature. Conversely, the thermally-equilibrated case is typified by: i) small S$_n$-S$_1$ (or T$_n$-T$_1$) energy gaps, so that thermal equilibrium between different ES can take place provided that enough thermal vibrational energy is available; and ii) the lack of excitation wavelength-dependency. TADF-phosphorescent dual emitters and other dual triplet-triplet emitters of, e.g., Ir(III)[12] fall in this second category. For the latter cases, the emission switch might be seen as a violation of Kasha-rule or not depending whether one considers the adiabatic energy difference between the optimized ES structures, or one examines the ES energy differences at their respective minima, as it often observed that the higher-lying ES becomes the lowest ES at the vicinity of its optimized geometry. Besides these conceptual problems, deciphering the actual origins of these anti-Kasha emission poses, in many cases, significant challenges both from the experimental and computational viewpoints, since the intricate photodeactivation dynamics is controlled by subtle variations of the electronic properties that are affected by both spin–orbit and vibronic interactions, as well as by the fine interplay between medium rigidity/polarity and temperature. Despite these difficulties, one of us reported in 2014 the first computational evidence of an anti-Kasha emission for a Pt(II) complex,[13] which has been followed by several additional contributions.[14,15,16]

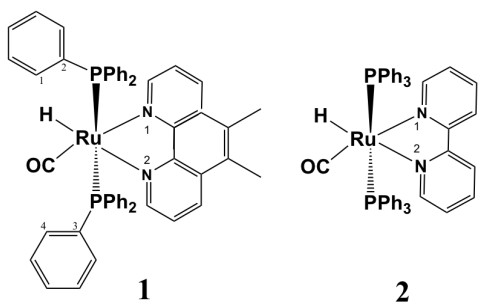

Scheme 1. Chemical structures of complexes **1-2**.

Recently, Kamecka et al. reported several heteroleptic [Ru(H)(CO)(N^N)(tpp)$_2$]$^+$ complexes,[17] some of them exhibiting temperature-dependent emissive properties, notably for the N^N=dmphen derivative (see **1** in Scheme 1). In contrast, the N^N=bpy derivative (**2**) did not show any $^3$MLCT to $^3$LC emission switch when cooling down the samples. Photon absorption is an ultrafast process, so that the orientation of the solvent molecules remains adapted to the charge distribution of the solute in its electronic ground state. Conversely, photon emission, which is a much slower process, implies a solute geometry relaxation in the ES as well as a solvent reorganization to adapt itself to the solute charge distribution in the electronic ES, before the actual photon emission takes place. At cryogenic temperatures, the rigidity of the media hinders, in principle, the solvent reorganization. Having in mind these facts, Kamecka and coworkers hypothesized that the $^3$MLCT state of **1** is the lowest-lying state at RT but not at 77 K, as the $^3$MLCT and $^3$LC states are differently affected by the lack of environmental relaxation at 77 K. However, they were not able to provide compelling proofs of this interesting hypothesis. In the following we address from a computational viewpoint these issues and we additionally explore the possibility of a thermally-equilibrated anti-Kasha scenario.

Relevant experimental data for **1-2** at 77K and RT are collected in Table 1. The experimental emission spectrum of **1** shows an unstructured band at RT peaking at 615 nm; whilst it shows a structured band peaking at 490 nm at 77K. At the latter temperature, a multicomponent decay, with different averaged lifetimes at the red and blue tails of the emission was recorded.[17] These observable data agree with a switch from a $^3$MLCT-like emission at RT to dual photoluminescence from both the $^3$LC and $^3$MLCT at 77K, as suggested by the experimental team. Conversely, an unstructured $^3$MLCT-like emission band is observed for **2** regardless of temperature. To get insights into the emissive characteristics of **1-2** we performed density functional theory (DFT) and time-dependent (TD)-DFT calculations (see Computational Details). More in details, the optimized geometries of different triplet excited states were determined; and subsequently the phosphorescence decay rates were evaluated at these geometries. Selected geometrical parameters of the optimized ground ($^1$GS) and relevant triplet state (i.e., $^3$MLCT, $^3$LC) minima are given in Table 2 for **1-2**. The assignment of the triplet character is based on the spin density distributions shown in Figure 1. Let us start our discussion with **1**. Its $^3$LC state is fully localized on the ancillary ligand (see Figure 1). Its optimized structure closely resembles the $^1$GS geometry (see Table 2). Both geometries bear a pseudo-octahedral environment around the Ru atom. In contrast, the $^3$MLCT geometry displays larger displacements with respect to the $^3$LC and $^1$GS geometries. These distortions include: (i) shorter Ru-N bond distances; (ii) asymmetric stretching of the Ru-P bonds; (iii) a smaller ∢$_{P-Ru-P}$ value; and (iv) the torsion of of one phenyl ligand (i.e., compare the dihedral angles ∢$_{C1-4}$ in Table 2). As shown in Figure 1, the spin density of the $^3$MLCT state is distributed among the Ru atom and the N^N ligand. A similar geometry and spin density distribution is obtained for the $^3$MLCT state of **2** (see Table 2 and Figure 1), which is the sole excited state involved in emission for **2**.

We now address the origins of phosphorescence. The phosphorescence emission maxima were computed on the basis of ΔSCF-PCM-B3LYP calculations, which are obtained as the energy difference between the triplet ES at its optimized geometry and the ground state at the same geometry; and through the calculation of vertical singlet-triplet PCM-TD-DFT excitations at their respective triplet optimized geometries. The results for **1-2** are listed in Table 3. Both approaches underestimate the position of the emission energy maxima with respect to the experimental evidences, especially in the case of the vertical singlet-triplet TD-DFT calculations. Provided that we did not explicitly include vibronic effects in these calculations, a reasonable agreement with the experimental emission spectra is nevertheless obtained (e.g., compare the ΔSCF-PCM-B3LYP computed value for **2** -1.57 eV- with the experimental one -1.97 eV-). For **1**, our calculations further support the nature of the ES involved in the experimentally-observed temperature-dependent emission switch. Hence, the ΔSCF-PCM-B3LYP computed emission maxima for the $^3$MLCT and $^3$LC states are 1.59 and 2.12 eV, respectively; in reasonable agreement with the blue-shift observed experimentally (see Table 1). To assess solvent relaxation effects, the perturbative corrected linear response (cLR)-PCM model[18] was used in combination with the TD-DFT calculations. In such a scheme, the variations in the solvent's cavity charges to the charge distribution in the electronic ES are taken into account using the one-particle TD-DFT density matrix. cLR-PCM-TD-DFT calculations are generally more accurate than their LR counterparts for emission energies.[19] The results are shown in Table 3. The cLR-PCM-TD-DFT values are slightly red-shifted (by ca. 0.2 and 0.3 eV) for the $^3$MLCT and $^3$LC emissions; respectively, but the energetic order is overall maintained with both solvation models. Additionally, the $^3$MLCT state is found to be always the lowest adiabatic ES regardless of the solvation scheme used. Therefore, the blue-shift observed experimentally at 77 K for **1** cannot be explained by a putative stabilization of the $^3$LC state at cryogenic temperatures, as suggested by Kamecka and coworkers. Indeed, the dipole moments of the ES and GS are rather similar (see Table 3); hinting at a limited impact of solvent relaxation effects.

Table 1. Experimental spectroscopic and photophysical properties of **1-2** at 77 K and RT. [a]

|   | $\lambda_{em}$ (77K) [nm, (eV)] | $\lambda_{em}$ (RT) [nm, (eV)] | $k_r$ (RT) [s$^{-1}$] | $\Phi_{em}$ (RT) |
|---|---|---|---|---|
| 1 | 490 (2.53), 525 (2.36), 561 (2.21) [b] | 615 (2.01) | 3.5×10$^3$ | 0.0012 |
| 2 | 551 (2.25) | 631 (1.97) | 5.8×10$^3$ | 0.0018 |

[a] From Ref. 17 (data in dichloromethane solutions at room temperature and methanol/ethanol 1:1 glasses at 77 K). [b] Vibrationally resolved peaks. The most intense band corresponds to the 0–1 transition.

Table 2. Relevant bond distances (Å) and angles (°) of **1-2** in their ground and triplet excited state minima.[a]

|   | Ru-N$_1$ / Ru-N$_2$ | Ru-H / Ru-C | Ru-P / Ru-P | $\alpha_{P-Ru-P}$ | $\alpha_{C1-4}$ | $\alpha_{H-C-P-P}$ |
|---|---|---|---|---|---|---|
| **1** – $^1$GS | 2.21 / 2.23 | 1.62 / 1.86 | 2.44 / 2.44 | 169.4 | 0.0 | -39.5 |
| **1** – $^3$LC | 2.20 / 2.21 | 1.63 / 1.86 | 2.44 / 2.44 | 169.2 | 3.9 | -40.0 |
| **1** – $^3$MLCT | 2.15 / 2.15 | 1.64 / 1.90 | 2.41 / 2.62 | 139.4 | 82.2 | 19.1 |
| **1** – TS | 2.21 / 2.21 | 1.63 / 1.86 | 2.44 / 2.46 | 165.8 | 51.1 | -38 |
| **2** – $^1$GS | 2.20 / 2.22 | 1.62 / 1.86 | 2.44 / 2.44 | 169.2 | 0.0 | -39.5 |
| **2** – $^3$MLCT | 2.15 / 2.14 | 1.64 / 1.90 | 2.41 / 2.54 | 139.7 | 85.2 | 19.7 |

[a] See atom labelling in Scheme 1.

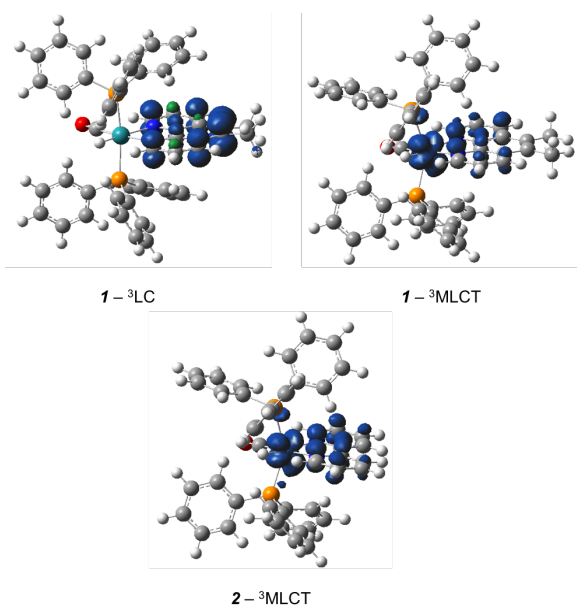

Figure 1. Spin density distributions (isoval = 0.006) for the lowest triplet excited states of **1-2**.

We now propose an explanation of the anomalous emissions observed for **1**. Its schematic Jablonski diagram is shown in Figure 2, which includes both $^3$MLCT and $^3$LC minima. The $^3$MLCT state is found to be the lowest adiabatic emissive state, i.e., the Kasha state, being 0.171 eV (4.0 kcal/mol) below the $^3$LC state. We additionally optimized the geometry of the transition state (TS) connecting both minima, and it was found 0.08 eV (1.8 kcal/mol) above the $^3$LC minimum. The TS optimized geometry resembles the one of $^3$LC (see Table 2). Its characteristic imaginary mode (31i cm$^{-1}$) mainly involves the torsion of the $v_{C1-4}$ angle and the bending of the $v_{P-Ru-P}$ angle. Thus, due to the steric hindrance between the two –PPh$_3$ units with decreasing $v_{P-Ru-P}$ angles, one of the phenyl rings is forced to twist in order to evolve to the $^3$MLCT minimum. As schematically shown in Figure 2, upon excitation to the singlet manifold (S$_n$), ultrafast ISC to the triplet manifold (T$_m$) will occur. The well of the $^3$LC minimum will preferentially be populated first after IC and vibrational relaxation, as it is closer to the Franck-Condon region. If enough thermal energy is available the system will be able to surmount the rather small $^3$MLCT→$^3$LC activation barrier and populate the well of the $^3$MLCT minimum in an irreversible or reversible manner depending on the given temperature. The kinetic model arising from such a scenario, including all possible ES decay channels is depicted in Scheme 2. The $^3$LC-$^3$MLCT equilibrium is controlled by the forward $k_+$(T) and reverse $k_-$(T) rate constants, which can be expressed in an Arrhenius form

$$k_{+/-}(T) = A\exp(-E_{act}/k_B T) \quad (1),$$

where $A$ is the preexponential factor, $E_{act}$ the corresponding activation energy for a forward or reverse process, and $k_B$ is the Boltzmann constant. At 77 K, in view of the experimental evidences, $^3$LC-like emission is predominantly observed (but coexisting with $^3$MLCT-like emission). Using the computed activation barriers (1.8 and 5.8 kcal/mol for the forward and reverse processes, respectively) a $k_+/k_-$ =2.3×10$^{11}$ value is obtained at 77 K; and consequently the $^3$LC→$^3$MLCT process is irreversible at this temperature (i.e., non-thermally equilibrated case). Therefore, at this limit condition (i.e., negligible $k_-$ value, since $k_+ >> k_-$) and according to the kinetic model depicted in Scheme 2, the quantum yields for the $^3$LC-like and the $^3$MLCT-like emissions can be expressed as

$$\Phi_{^3LC}(77\ K) = \frac{k_{phos1}}{k_{phos1}+k_{nr1}+k_+(77\ K)} \quad (2),$$

$$\Phi_{^3MLCT}(77\ K) = \frac{k_{phos2}k_+(77\ K)}{(k_+(77\ K)+k_{phos1}+k_{nr1})(k_{phos2}+k_{nr2})} \quad (3),$$

respectively. $^3$LC-like emission is observed because $k_{phos1}$ competes with $k_{nr1}$ and $k_+$ (77 K). Additionally, the $^3$MLCT-like emission is concomitantly observed due to the confluence of two factors: i) $k_{phos2} >> k_{phos1}$ (see the computed radiative rates in Table 3 and the discussion below); and ii) a non-negligible $k_+$ (77 K) value. Let us now discuss the RT scenario. A $k_+/k_-$ = 863 value is obtained at RT, which is much smaller than at 77 K, and thus, this implies that a thermal equilibrium between the

$^3$LC and $^3$MLCT states is likely. Consequently, equations (2) and (3) do not hold anymore.

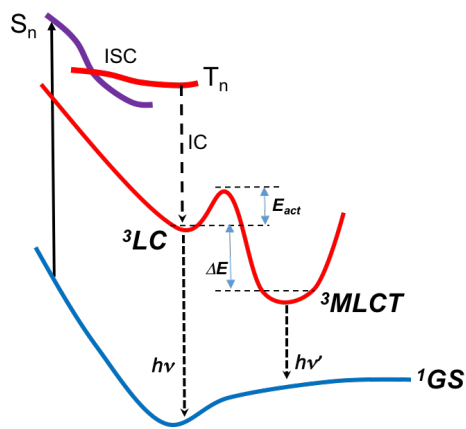

Figure 2. Schematic Jablonski diagram of **1**, including the triplet excited states involved in emission.

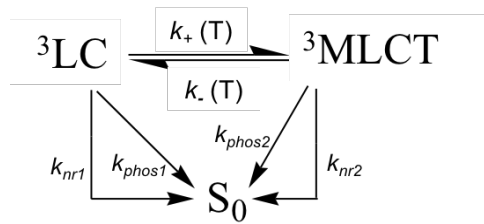

Scheme 2. Kinetic model including all possible ES decay channels for **1**, including the temperature-dependent forward $k_+$(T) and reverse $k_-$(T) decay rates, the radiative rates ($k_{phos1}$ and $k_{phos2}$), and the nonradiative decay rates ($k_{nr1}$ and $k_{nr2}$). Note that the nonradiative decay rates may involve different processes.

In such a thermally-equilibrated scenario, we can assess whether dual photoluminescence, or solely emission from the higher or lower ES will arise with the following expression,[5]

$$\frac{\phi(^3MLCT)}{\phi(^3LC)} = \frac{k_r(^3MLCT)}{k_r(^3LC)} e^{\left[\frac{-\Delta E(^3LC - {}^3MLCT)}{k_B T}\right]}, \quad (4)$$

where the $^3$MLCT/$^3$LC ratio of quantum yields is determined by the ratio of their radiative rates, the ΔE value (note that for a thermally-equilibrated scenario, the use of ΔE instead of $E_{act}$ is best suited), and the temperature T. Generally, if the ratio is <10$^{-2}$, only $^3$LC emission will be observed. Emission from the $^3$MLCT state will be predominant if the ratio is >10$^{+2}$ and; finally, values between these two thresholds lead to dual emissions. The basic approximation behind Eq. (4) is the assumption that $k_{nr1} \approx k_{nr2}$. We underline that the calculation of $k_{nr}$ poses significant challenges from a computational viewpoint, as often there are multiple channels for nonradiative deactivation in organometallic complexes.[20,21] In order to assess the temperature-dependent photoluminescence properties for **1**, we evaluated the $k_r$ values arising from both the $^3$MLCT and $^3$LC ES with quadratic-response (QR) TD-DFT calculations (see Computational Details). The results are shown in Table 3. The computed rates are slightly underestimated with respect to their experimental ones, as commonly found for perturbative approaches as compared to self-consistent ones.[5] The ratio computed with Eq. (4) at this temperature is well above >10$^{+2}$, pointing to sole $^3$MLCT emission, as the radiative decay rate from this state is two orders of magnitude larger than from the $^3$LC state (see Table 3); and in addition is the lowest energy ES. Therefore, the switch from dual photoluminescence at 77 K to sole $^3$MLCT-like emission at RT roots on the feasibility to attain the thermal equilibrium between both ES.

Table 3. Computed dipole moments of the emissive states, emission maxima and $k_r$ values from the emissive states of **1**-**2**.

| | μ ($\mu_x$; $\mu_y$; $\mu_z$) [Total] Debye | ΔSCF-PCM-B3LYP nm, (eV) | PCM-TD-B3LYP CH$_2$Cl$_2$ // EtOH *cLR-PCM-TD-B3LYP* CH$_2$Cl$_2$ // EtOH nm, (eV) | $k_r$ s$^{-1}$ |
|---|---|---|---|---|
| 1– $^1$GS | (0.0; 8.5; 1.8) [8.69] | - | - | - |
| 1– $^3$MLCT | (1.9; -0.6; 1.9) [2.74] | 779 (1.59) [a] | 816 (1.52) // 804 (1.54) *919 (1.35) // 932 (1.33)* | 7.6 × 10$^2$ |
| 1– $^3$LC | (-0.3; 9.5; 1.7) [9.67] | 585 (2.12) [b] | 422 (2.94) // 418 (2.97) *470 (2.64) // 479 (2.59)* | 2.2 |
| 2– $^1$GS | (0.0; 7.9; 4.1) [8.87] | . | - | . |
| 2– $^3$MLCT | (-1.3; 1.5; 1.4) [2.42] | 792 (1.57) [a] | 836 (1.48) // 822 (1.51) *990 (1.25) // 967 (1.28)* | 2.1 × 10$^2$ |

[a] Computed in CH$_2$Cl$_2$. [b] Computed in ethanol.

## Conclusions

In conclusion, we have unraveled from first principles the temperature-dependent photoluminescent properties of several Ru(H)(CO)(N^N)(tpp)$_2$]$^+$ complexes. The anomalous emissions arise from a kinetically controlled scenario. At RT thermal-equilibration between the $^3$MLCT and $^3$LC states is possible, leading to sole emission from the $^3$MLCT state. Thermal equilibrium is not possible at 77 K, leading to anomalous (or anti-Kasha depending on how we define the violation of Kasha's rule) emissions from the $^3$LC state, which coexist with $^3$MLCT emissions. We additionally rule out other possibilities for the origins of these anomalous emissions, such as different ES solvent relaxation effects. The deep understanding of the temperature-dependent photoluminescent properties is of pivotal importance for the design of tailored materials for optoelectronics, including ES managers of hot excitons. Towards the latter aim, anti-Kasha emitters are promising strategies to be further explored in the community.

## Acknowledgements

DE thanks funding from the European Union's Horizon 2020 research and innovation programme under the Marie Sklodowska-Curie grant agreement No 700961. This work used computational resources of the CCIPL and the IDRIS/CINES.

## Computational Details

All calculations for **1-2** are based on density functional theory (DFT). The geometries of the $^1$GS and of the lowest triplet ES ($^3$MLCT and $^3$LC), as well as of the transition state (TS) were optimized using the B3LYP hybrid exchange-correlation functional[22,23] in combination with the 6-31G* atomic basis set for all atoms but the metal centers. Relativistic effects were included for the Ru atom by using the LANL2DZ pseudopotential.[24] The nature of the stationary points was confirmed by computing the Hessian at the same level of theory. Towards the exploration of the triplet potential energy surfaces we pre-optimized the first five triplet excited states at the TD-B3LYP level of theory. These optimized geometries were used as initial guess for the final PCM-UB3LYP optimizations leading to $^3$MLCT and $^3$LC for **1** and $^3$MLCT for **2**. The phosphorescence emission spectra were simulated on the basis of ΔSCF-B3LYP calculations and; additionally, singlet-triplet vertical TD-B3LYP excitations were performed at the triplet optimized geometries using the same basis sets as in the optimizations. The phosphorescence calculations and the optimizations of the lowest triplet ES took into account solvation (see text for solvent used) effects within the PCM framework.[25] We have tested both the linear-response and the corrected linear-response variants of the PCM for excited-state properties, considering the non-equilibrium regime in both cases. All these calculations were carried out with the Gaussian09 program package.[26] The phosphorescence radiative decay rates were computed using the QR TD-B3LYP approach,[27] as implemented in the Dalton program,[28] at the optimized geometries of the different emissive triplet states. For the sake of computational ease in the latter calculations the phenyl groups were replaced by methyl groups. Towards this aim, the TD-B3LYP excitations leading to the $^3$MLCT and $^3$LC states were first assigned. In the QR TD-B3LYP calculations the 6-31G and raf-r basis set were used for light atoms and Ru centers, respectively. Scalar relativistic effects were included with the Douglas-Kroll-Hess second order (DKH2) Hamiltonian.[29]

## Notes and references

**TOC**

Thermal equilibration between excited states or solvent effects: unveiling the origins of anti-Kasha emissions in heteroleptic [Ru(H)(CO)(N^N)(tpp)$_2$]$^+$ complexes.

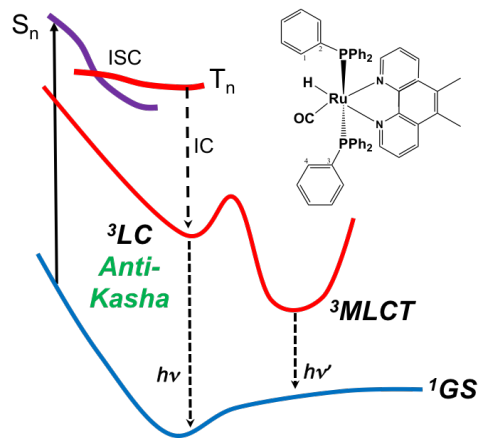